\documentstyle[emulateapj,epsf]{article}
\submitted{ApJ, in press}

\def\sun{\hbox{$\odot$}}

\def\lesssim{\mathrel{\hbox{\rlap{\hbox{\lower4pt\hbox{$\sim$}}}\hbox{$<$}}}}
\def\gtrsim{\mathrel{\hbox{\rlap{\hbox{\lower4pt\hbox{$\sim$}}}\hbox{$>$}}}}

\def\arcdeg{\hbox{$^\circ$}}

\let\la=\lesssim                        
\let\ga=\gtrsim
\newcommand{\mm}[1]{\mbox{$#1$}}
\newcommand{\unit}[1]{\ifmmode \:\mbox{\rm #1}\else \mbox{#1}\fi}

\newcommand{\expec}[1]{\mm{\left\langle #1 \right\rangle}}
\newcommand{\mone}{\mm{^{-1}}}

\newcommand{\kms}{\unit{km~s\mone}}
\newcommand{\kpc}{\unit{kpc}}
\newcommand{\mpc}{\unit{Mpc}}
\newcommand{\hkpc}{\mm{h\mone}\kpc}
\newcommand{\hmpc}{\mm{h\mone}\mpc}

\newcommand{\secref}[1]{\S\ref{sec:#1}}
\newcommand{\brref}[1]{(equation~\ref{eq:#1})}
\newcommand{\eqref}[1]{equation~(\ref{eq:#1})}

\newcommand{\figref}[1]{Fig.~\ref{fig:#1}}

\newcommand{\old}[1]{}

\lefthead{HUDSON ET AL.} 
\righthead{GALAXY--GALAXY LENSING IN THE HDF} 

\begin{document} 
\title{Galaxy--Galaxy Lensing in the Hubble Deep Field: \\ The Halo
  Tully-Fisher Relation at Intermediate Redshift}

\author{Michael J. Hudson\altaffilmark{1} \& Stephen D. J. Gwyn}
\affil{ Department of Physics \& Astronomy, University of Victoria,\\
  P.O. Box 3055, Victoria, B.C. V8W 3P6, Canada\\
  E-mail: hudson, gwyn@uvastro.phys.uvic.ca}
\altaffiltext{1}{CITA National Fellow}

\and 

\author{H{\aa}kon Dahle \& Nick Kaiser} 
\affil{Institute for Astronomy, 
2680 Woodlawn Drive, Honolulu, HI 96822, USA\\
E-mail: dahle, kaiser@ifa.hawaii.edu}

\begin{abstract} 
  A tangential distortion of background source galaxies around
  foreground lens galaxies in the Hubble Deep Field is detected at the
  99.3\% confidence level.  An important element of our analysis is
  the use of photometric redshifts to determine distances of lens and
  source galaxies and rest-frame $B$-band luminosities of the lens
  galaxies.  The lens galaxy halos obey a Tully--Fisher relation
  between halo circular velocity and luminosity; the typical lens
  galaxy, at a redshift $z = 0.6$, has a circular velocity of
  $210\pm40 \kms$ at $M_{B} = -18.5$, if $q_0 = 0.5$.

  Control tests, in which lens and source positions and source
  ellipticities are randomized, confirm the significance level of the
  detection quoted above.  Furthermore, a marginal signal is also
  detected from an independent, fainter sample of source galaxies
  without photometric redshifts.  Potential systematic effects, such
  as contamination by aligned satellite galaxies, the distortion of
  source shapes by the light of the foreground galaxies, PSF
  anisotropies, and contributions from mass distributed on the scale
  of galaxy groups are shown to be negligible.

  A comparison of our result with the local Tully--Fisher relation
  indicates that intermediate-redshift galaxies are fainter than local
  spirals by $1.0\pm0.6$ $B$ mag.\ at a fixed circular velocity.  This
  is consistent with some spectroscopic studies of the rotation curves
  of intermediate-redshift galaxies.  This result suggests that the
  strong increase in the global luminosity density with redshift is
  dominated by evolution in the galaxy number density.
\end{abstract} 


\section{Introduction} 
\label{sec:intro} 

The existence of dark matter halos around individual galaxies is now
well established, but a detailed understanding of the distribution of
the dark matter remains elusive.  Traditional probes of the galaxy
dark matter halos have been dynamical: from disk rotation curves on
scales of tens of kpc, to the dynamics of globular clusters, planetary
nebulae, satellites and companion galaxies on larger scales.  An
alternative probe of dark matter halos is the gravitational lens
distortion of the shapes of background ``source'' galaxies.  A great
advantage of the lensing approach is that no assumptions about the
dynamical state of the system are necessary.  Clusters of galaxies
have traditionally been the primary target of weak lensing studies.
Galaxies are much less massive than rich clusters, and so the lensing
signal is correspondingly weaker.  However, the large number of
lens--source pairs compensates, allowing a statistical detection of
the signal. After a null detection (Tyson et al.\ 1984), the subject
of galaxy-galaxy lensing remained dormant until the recent work of
Brainerd, Blandford \& Smail (1996, hereafter BBS).  BBS analyzed 90
square arcmin of deep two-color ground-based images and detected
galaxy--galaxy lensing from halos with typical circular velocities of
220 \kms.  They studied lens--source pairs with separations of 5 to 34
arcseconds, corresponding roughly to 20 to 125 \hkpc\ at the typical
lens redshift.  Recently, Dell'Antonio \& Tyson (1996, hereafter DT)
analyzed the Hubble Deep Field (Williams et al.\ 1996, hereafter HDF)
and detected galaxy--galaxy lensing within 5 arcseconds of the lens
galaxy, corresponding to approximately 16 \hkpc\ at their assumed mean
lens redshift.

In this paper, we investigate galaxy--galaxy lensing in the HDF.  Our
analysis differs significantly from that of DT, as will be shown
below.  A key feature of our analysis is the use of photometric
redshifts derived from the four HDF colors.  In \secref{data}, we
describe the lens and source catalogs and the photometric redshifts.
In \secref{method}, we present our maximum-likelihood method, and test
it with Monte Carlo simulations in \secref{monte}.  Our main results
are presented in \secref{results}.  The characteristics of the lens
sample are investigated in \secref{new}, and some potential
systematics are examined in \secref{sys}.  Finally, in
\secref{discuss}, we compare our results to other galaxy--galaxy
lensing studies and to Tully--Fisher (hereafter TF) results from disk
rotation curves at low and intermediate redshifts.  We assume a
cosmology with $q_0 = 0.5$ and $H_0 = 100\,h\, \kms \mpc\mone$.

\pagebreak
\section{Data} 
\label{sec:data} 

Our galaxy catalogs are based on the wide field chips (WF2--4) of the
Hubble Deep Field version 2 `drizzled' images (Williams et al.\ 1996).
As a shorthand for the F300W, F450W, F606W and F814W passbands we use
$U_{\rm ST}$, $B_{\rm ST}$, $R_{\rm ST}$ and $I_{\rm ST}$
respectively.  These magnitudes have the $ST$ system zero-point.

\subsection{Object Detection and Shape Parameters}

Object catalogs were extracted using two different software
packages: SExtractor version 1.0a (Bertin \& Arnouts 1996) and the
imcat software 
\footnote{%
  Documentation for the imcat software can be found at
  http://www-nk.ifa.hawaii.edu/$\sim$kaiser/imcatdoc/mainindex.html 
}
of Kaiser, Squires \& Broadhurst (1995, hereafter KSB).
SExtractor is optimized for photometry so it was used to generate the
catalog of lens galaxies.  The catalog of sources was generated
with imcat, which is optimized for extracting shapes of faint galaxies
and other parameters relevant to lensing.

In order to avoid the edges of the chips, where image quality is poor,
for both SExtractor and imcat catalogs we use only objects within the
following regions: (x,y) = (182,120) to (2006,1944) of WF2; (142,192)
to (1942,1992) of WF3; and (208,208) to (1976,1976) of WF4.

The lens galaxies were detected on the unsmoothed $I_{\rm ST}$
drizzled images.  The isophotal threshold was set to twice the r.m.s.\ 
sky (corresponding to $26.65 \, I_{\rm ST}\, {\rm mag}/{\rm square\,
  arcsecond}$), and a minimum of 25 pixels were required for inclusion
in the catalog.
We then use the object masks generated by
SExtractor to measure the fluxes within the same isophotal area in all
four of the HDF passbands.  

The source catalog was obtained using the imcat software on the
deeper $R_{\rm ST}$ images, using the object detection algorithm
described by KSB.  This basically consists of smoothing the image with
a set of ``Mexican hat'' type filters with a range of widths,
detecting peaks in the smoothed images and assigning a detection
significance $\nu$ to each peak.  The smoothing filter width, $r_{g}$,
which gave maximum $\nu$ was found for each individual object. For our
source catalog, we excluded objects with $\nu < 10\sigma$, but the
true significance threshold was somewhat lower than that, since the
pixel-to-pixel noise is correlated in the drizzled HDF images. We also
excluded objects with half-light radius below 1.3 pixels.

For each object, we measure weighted quadrupole moments of the
intensity distribution $f(\vec{\theta})$,
  \begin{equation}
    I_{ij} = \int d^{2}\theta W(\theta) \theta_{i} \theta_{j}
    f(\vec{\theta}){\rm ,}
  \end{equation}
  where $W(\theta)$ was taken to be
  \begin{equation}
    W(\theta) = \exp(-\theta^{2} / 2 r_{g}^{2})\,.
  \end{equation}
From the quadrupole moments, we obtain the ellipticity polar%
\footnote{Polars, unlike vectors, rotate by $2\phi$ under a $\phi$
  rotation of the coordinate axes.}, $e_{\alpha} = \{ (I_{11} -
I_{22}) , 2I_{12} \} /(I_{11} + I_{22})$.

The susceptibility of change of $e_{\alpha}$ of a source due to a
shear $\gamma$ is
\begin{equation}
e'_{\alpha} -  e_{\alpha}  \equiv P^{\gamma}_{\alpha \beta} \gamma_{\beta}\,,
\end{equation}
where the prime denotes the ellipticity in the presence of a shear.
In the weak lensing regime ($\gamma \ll 1$), an idealized circular
source with unweighted ($W = 1$) second moments would have
$P^{\gamma}_{\alpha \beta} = 2 \delta_{\alpha \beta}$.  In practice,
however, the radial weighting and the seeing both complicate this
expression.  Fortunately, the correct expression for $P^{\gamma}$ is
readily obtained from the observables.  Luppino \& Kaiser (1997) give
the ``pre-seeing shear polarizability'' as
  \begin{equation}
    P^{\gamma}_{\alpha \beta} = P_{\alpha \beta}^{\rm sh} - P_{\alpha
      \beta}^{\rm sm} P_{\alpha \beta}^{\rm sh}(*)/P_{\alpha
      \beta}^{\rm sm}(*) {\rm ,}
  \end{equation}
where $P_{\alpha \beta}^{\rm sh}$ and $P_{\alpha \beta}^{\rm sm}$
are the post-seeing shear polarizability and smear polarizability
matrices, as defined by KSB, but correcting for a minor error in
their eqs.\ A12 and B12 (Hoekstra et al.\ 1997). 
The asterisks denote average values for stellar objects in the field.
Note that this relation is only strictly valid for a Gaussian PSF, but
the drizzled HDF PSFs are fairly close to Gaussian, so we chose to
adopt this relation.

In practice, however, we ended up using ``scalar'' $P^{\gamma}$
values, 
\begin{equation}
  P^{\gamma} =\frac{1}{2} \left(P_{\alpha \alpha}^{\rm sh} - P_{\alpha
      \alpha}^{\rm sm} \frac{P_{\alpha \alpha}^{\rm sh}(*)}{P_{\alpha
      \alpha}^{\rm sm}(*)} \right) ,
\end{equation}
where 
repeated subscripts denote the trace of the matrix, as these were
found to be slightly less noisy than using the full
$P^{\gamma}_{\alpha \beta}$ matrices for each individual source. We
excluded sources with $P^{\gamma}$ below 0, since these had
ill-defined shape parameters.  This left us with a catalog of 1622
objects.

In order to model the ellipticity distribution, we have experimented
with two functional forms.  The first model is just a two-dimensional
Gaussian, in which the probability density distribution of each
component of the ellipticity polar is given by
\begin{equation}
  P(e)de \propto
  \exp\left({-\frac{e^2}{2\sigma_e^2}}\right)\,de \,.
\label{eq:gaussellip}
\end{equation}
with $e \le 1$.  For the source catalog, the dispersion in ellipticity
is $\sigma_e = 0.185$, independent of magnitude.  The histogram of
ellipticities (with both $e_1$ and $e_2$ components stacked together)
is shown in \figref{ellhist}.  The Gaussian model is shown by the
dotted line.  This model is a good fit to the tails of the
distribution, but it underestimates the number of galaxies with small
ellipticity.

\vbox{%
\begin{center}
\leavevmode
\hbox{%
\epsfxsize=8.9cm
\epsffile{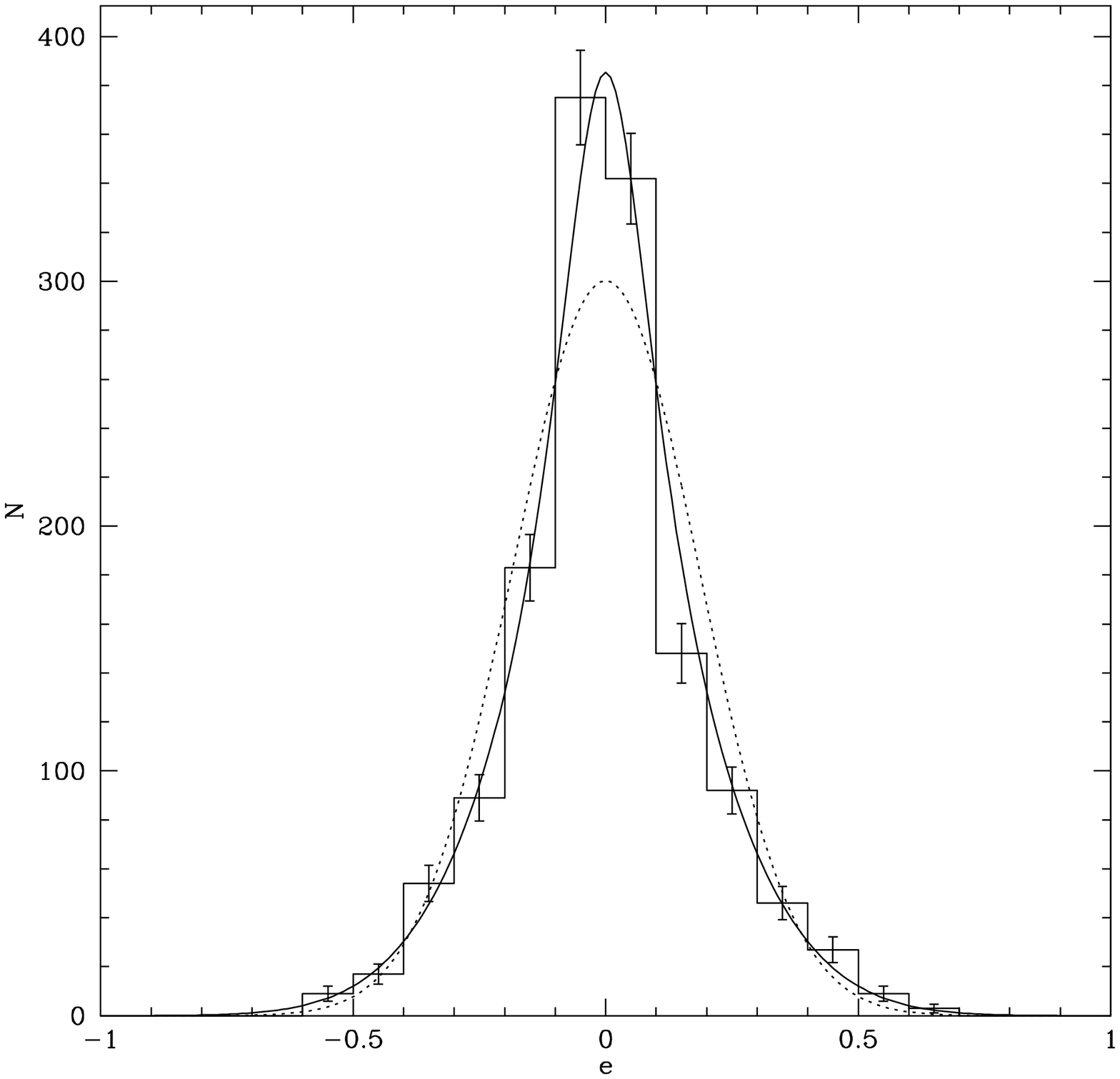}}
\begin{small}
\figcaption{%
  Histograms of ellipticity of source galaxies.  Both $e_1$ and $e_2$
  components are stacked together.  The dotted curve shows 
  the Gaussian of best fit, whereas the solid curve shows the
  modified Gaussian (equation [\ref{eq:gausslorellip}]) of best fit. 
\label{fig:ellhist}}
\end{small}
\end{center}}

After some trial and error, we have found a good fit using modified
Gaussian model with the functional form
\begin{eqnarray}
  P(e)de & \propto &
  \exp\left({-\frac{e^2}{2 \sigma_e^{'2}}}\right) \times
  \nonumber \\ & & \left( 1 + \frac{1}{\pi a_e}
  \frac{1}{(1+(e/a_e)^2)}\right)\,de
\label{eq:gausslorellip}
\end{eqnarray}
with $\sigma'_e = 0.23$, and $a_e = 0.13$.  This fit is shown by the
solid line in \figref{ellhist}.  We adopt the Gaussian fit of
\eqref{gaussellip} as the default in this paper, but we also consider
results from the modified Gaussian.

Note that we have measured these ellipticities from the observed
galaxy ellipticities, {\em after\/} they have been lensed.  In
principle, the correct procedure would be to use the ellipticities
before lensing.  In practice, however, in our model the typical shear
$\gamma \ll \sigma_e$, so subtracting this shear off in quadrature
makes no difference to the values derived above.

\subsection{Photometric Redshifts}

For every object in the lens and source catalogs, we obtain
photometric redshifts as described in Gwyn \& Hartwick (1996).  These
photometric redshifts are based on 0.2 arcsecond radius aperture
photometry.  The method essentially consists of converting these
magnitudes to a low-resolution spectral energy distribution (SED).  A
set of template spectra of all Hubble types and redshifts ranging from
$z = 0$ to $z = 5$ is compiled. The redshifted template spectra are
reduced to the passband averaged fluxes at the central wavelengths of
the observed passbands.  The SED derived from the observed magnitudes
of each object is compared to each template spectrum in turn. The best
matching spectrum, and hence the redshift, is then determined by
minimizing $\chi^2$.  Some improvements have been made to the method
of Gwyn \& Hartwick (1996): the intergalactic Lyman blanketing
corrections of Madau (1995) are now incorporated; and for low
redshifts ($z < 1.5$) the empirical SEDs of Coleman, Wu \& Weedman
(1980), as extended by Ferguson \& McGaugh (1995), are used in
preference over the evolving SEDs of Bruzual \& Charlot (1993).  Both
of these corrections are found to improve the accuracy of the
photometric redshifts when compared to spectroscopic redshifts.

We limit both lens and source catalogs to galaxies with isophotal
$I_{\rm ST}$ magnitudes less than 28.  If a galaxy has a spectroscopic
redshift, we use this in preference to the photometric redshift.  From
a comparison of spectroscopic and photometric redshifts, it is found
that at low redshift ($z<1.5$) the $1\sigma$ uncertainty in $z$ is
0.15, at high redshift ($z>1.5$) the uncertainty is 0.35.  In a recent
independent test, Hogg et al.\ (1998) obtained spectroscopic redshifts
of HDF galaxies with $z < 1.5$ and solicited photometric redshift
predictions prior to publication of the spectroscopic data.  The
Victoria photometric redshifts were found to be accurate to $|\Delta
z| < 0.3\,(0.1)$, 19 (15) times out of 19, when compared to
high-quality spectroscopic redshifts. This is in good agreement with
the errors quoted above.

For the lens catalog, we then use the redshifts to obtain absolute
isophotal magnitudes in the rest-frame $B$-band (on the Johnson
magnitude system). We wish to avoid the uncertainties of
$k$-corrections, so for each lens galaxy we determine its rest-frame
$B$-band magnitude by interpolation from its measured HDF magnitudes.
This limits the lens catalog to the redshift range $0 \le z \le 0.85$.
The lens catalog contains 208 galaxies.  The source catalog consists
of 697 galaxies with photometric redshifts, and includes the lens
catalog as a subset.  The top panel of \figref{zhist} shows the
redshift distribution of the source catalog.

\vbox{%
\begin{center}
\leavevmode
\hbox{%
\epsfxsize=8.9cm
\epsffile{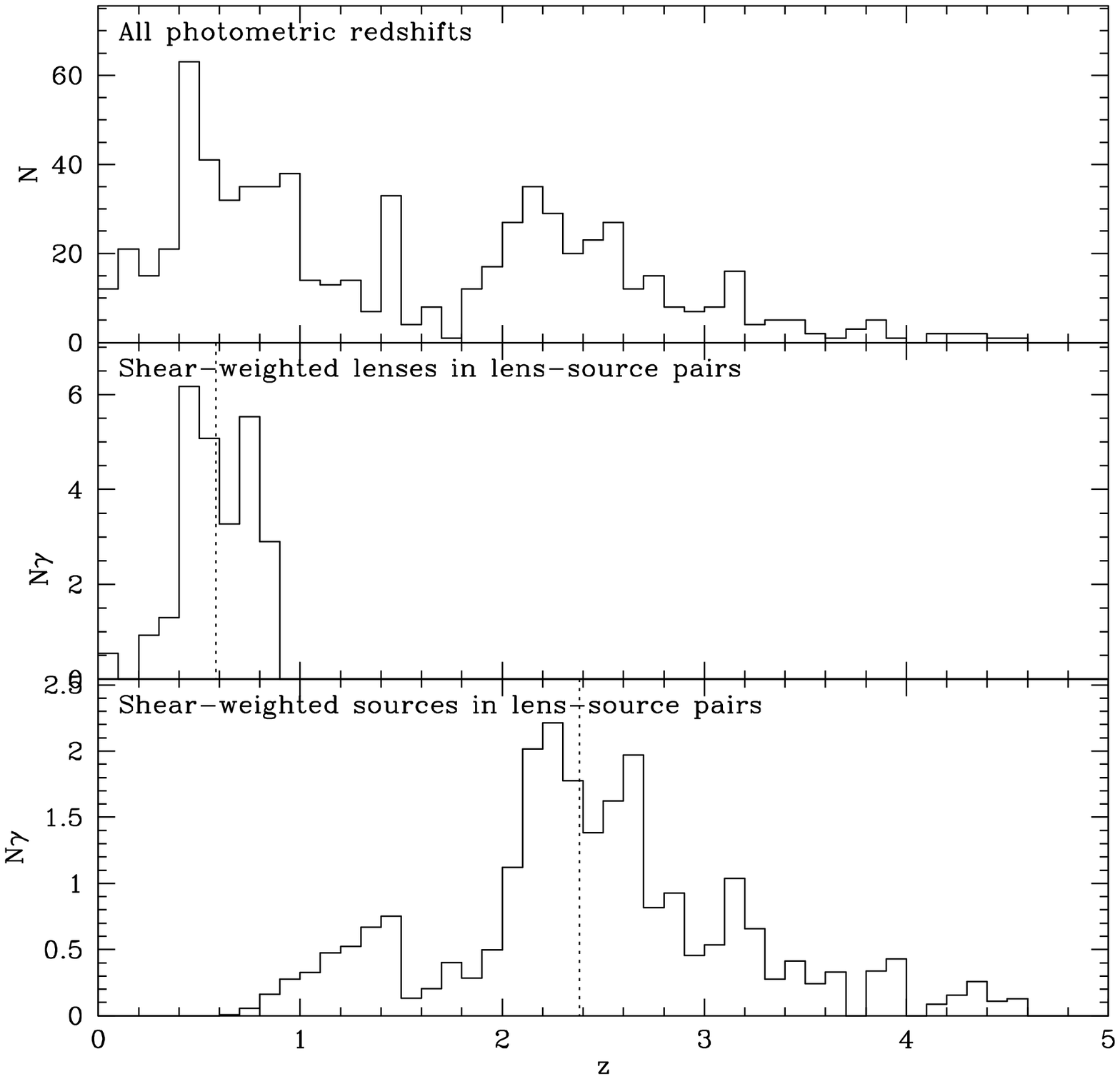}}
\begin{small}
\figcaption{%
  Redshift histograms.  The top panel shows redshift histogram of the
  sample.  The middle and bottom panels show the shear-weighted
  histogram of lens and sources respectively in all lens--source
  pairs (see \secref{new} for a description of shear-weighting).  The
  means of the weighted distributions are shown by the
  dotted vertical lines.
\label{fig:zhist}}
\end{small}
\end{center}}

\section{Method}
\label{sec:method}

In order to determine the masses of the halos, we perform a maximum
likelihood analysis similar to that of Schneider \& Rix (1997).  The
outline of this method is as follows: for a given model of the lens
galaxy mass density profiles, calculate the shear at every source
galaxy due to lensing by all galaxies in its foreground, thus obtain
the true unlensed ellipticity. Then calculate the likelihood of this
ellipticity given the ellipticity distribution determined above.  A
sum over all source galaxies yields a log likelihood which can be
minimized with respect to the parameters of the mass model.

Our model for the galaxy dark matter halos follows BBS. Halos are
assumed to be spherical singular truncated isothermal spheres with
truncation radius, $s$, with density
\begin{equation}
  \rho(r) = \frac{V^2 s^2}{4 \pi G r^2 (r^2 +s^2)}\,,
  \label{eq:stis}
\end{equation}
where $V$ is the circular velocity in the limit $r \ll s$ and $G$ is
Newton's constant.

It is convenient to introduce the Einstein radius for a pure
isothermal sphere,
\begin{eqnarray}
  \theta_{\rm E} & = & 2 \pi \left(\frac{V}{c}\right)^2 \frac{D_{\rm
      ls}}{D_{\rm s}} \\ 
     & = & 0.577 \left( \frac{V}{200
    \kms}\right)^2 \left(\frac{D_{\rm ls}}{D_{\rm s}}\right) {\rm
    arcseconds}
\end{eqnarray} 
where $D_{\rm s}$ and $D_{\rm ls}$ are the angular diameter distances
to the source and between lens and source.  At typical lens and source
redshifts ($z_{\rm l} = 0.6$ and $z_{\rm s} = 2.4$), $D_{\rm
  ls}/D_{\rm s} = 0.54$. For a pure isothermal sphere, the tangential
shear at a separation $\theta$ is
\begin{equation}
  \gamma_t = \frac{\theta_{\rm E}}{2 \theta}\,.
  \label{eq:gammatiso}
\end{equation}
For the truncated isothermal sphere, the tangential shear, $\gamma_t$
is given by (BBS):
\begin{equation}
  \gamma_t = \gamma_0 (V, s, z_{\rm l}, z_{\rm s}) \,
  G(\theta/\theta_s)\,,
  \label{eq:gammat}
\end{equation}
where $\gamma_0 = \theta_{\rm E}/(2 \theta_s)$, $\theta_s = s/D_{\rm
  l}$ is the angle subtended by $s$, $D_{\rm l}$ is the angular
diameter distances to the lens, and
\begin{equation}
  G(X) = \frac{(2+X)(1+X^2)^{1/2}-(2+X^2)}{X^2(1+X^2)^{1/2}}
  \label{g} \,.
\end{equation}
For $X \ll 1$, $G(X) \sim 1/X$, whereas for $X \gg 1$, $G(X) \sim
2/X^2$.

The shear polar, $\gamma_{\alpha}$, is then $\{-\gamma_t\, \cos(2\phi),
  -\gamma_t\, \sin(2\phi)\}$ where $\phi$ is the position angle of the
line joining the lens and source galaxies.  The negative sign accounts
for the fact that the shear is in the tangential direction.

For each lens galaxy, we obtain the circular velocities, $V$, by
assuming that the dark matter halos follow a Tully--Fisher relation
of the form
\begin{equation}
  V= V_{\rm f} \left(\frac{L}{L_{\rm f}(z)}\right)^{\eta}\,,
  \label{eq:tf}
\end{equation} 
where $V_{\rm f}$ is the circular velocity of a galaxy with fiducial
luminosity $L_{\rm f}(z)$, and $\eta$ is the slope of the relation.
We allow for luminosity evolution of the lenses by assuming that the
fiducial luminosity scales with redshift as
\begin{equation}
  L_{\rm f}(z) = L_{\rm f}(z=0.6) \left(\frac{1+z}{1.6}\right)^{\zeta}\,,
\label{eq:evol}
\end{equation}
where $L_{\rm f}(z=0.6)$ is the fiducial luminosity of the
Tully-Fisher relation at $z = 0.6$, which is close to the effective
mean redshift of the lenses.  We adopt $M_{\rm f}(z=0.6) = -18.5$,
close to the effective mean magnitude of the lenses (see
\secref{new} below).

Finally, following BBS, we scale the truncation radius $s = s_{200}
(V/200 \kms)^2$.  However, our results below will show that $s_{200}$
is not well constrained.  We have also experimented with a linear
scaling relation ($s \propto V$) but find that the resulting $V_{\rm
  f}$ differs by less than 2 per cent.

For all lens-source pairs used in the analysis, the predicted shear
$\gamma \ll 1$, and is typically $\sim 0.02$.  In the weak shear
limit, it is valid to calculate the total predicted shear, $\gamma_j$,
at a source galaxy $j$ via a linear sum over the contributions
$\gamma_{ij}$ due to all foreground galaxies $i$:
\begin{equation}
  \gamma_j = \sum_i \gamma_{ij}(R_{ij}, V_i, s_i, z_i, z_j)
  \label{eq:sum}
\end{equation}
where the sum extends over all galaxies with $z_i < z_j$.  In
practice, we use only those pairs with a minimum redshift separation
of 0.5 and projected separations at the lens $R > 10$ h$^{-1}$ kpc%
\footnote{Note that a given source galaxy is not excluded from the
  analysis if it passes within less than 10 h$^{-1}$ kpc of a lens
  galaxy. However, the shear for that particular lens-source pair is
  not included in \eqref{sum}.  Note that the position angle of such
  close pairs is uncorrelated with the position angles of the pairs
  which are included in the sum, so that no bias arises from excluding
  such pairs.  The shear due to the close encounter will increase the
  scatter in ellipticity of such sources. This shear is negligible in
  comparison to the intrinsic scatter unless the encounter is within a
  few Einstein radii, $\sim 0.75$ arcseconds or $\sim 3 \hkpc$ for a
  typical lens.  We estimate that only a few per cent of sources are
  significantly affected in this way.}. Finally, we limit the sample
to pairs with an angular separations less than 30 arcseconds, so that
most pairs fall within the same WF chip.  The motivation for the
latter cut is discussed further in \secref{psf}.  There are 10156 such
lens-source pairs.

The ellipticity of a given source galaxy in the absence of lensing is
just $e_j = e'_j - \gamma_j P^{\gamma}_{j}$.  The likelihood is then
simply the probability of the corrected (unlensed) ellipticity given
the measured distribution of ellipticities of galaxies in the HDF.

For the Gaussian ellipticity distribution \brref{gaussellip}, a simple
sum over all source galaxies yields a log likelihood
\begin{equation}
  \log {\cal L} = \sum_j \left(-\frac{|e'_j - \gamma_j
    P^{\gamma}_j|^2} {2\sigma_{e}^2}\right)
\end{equation}
which is maximized by varying the free parameters $V_{\rm f}$,
$s_{200}$, $\eta$ and $\zeta$.  The quantity $2\,(\log {\cal L}_{\rm
  max} - \log {\cal L})$ has a $\chi^2$ distribution with the number
of degrees of freedom equal to the number of free model parameters. We
can evaluate the significance of the galaxy--galaxy lensing signal by
comparing the likelihood with that obtained by setting $V = 0$:
\begin{equation}
  \chi^2 = 2\,(\log {\cal L}_{\rm max} - \log {\cal L}_{V = 0}) \,.
  \label{eq:chi}
\end{equation}

Note that the shear for any one lens--source pair is small compared to
the intrinsic scatter in galaxy shapes. As a result, the lensing
signal is an average over many pairs; no single lens-source pair
dominates.  In essence, our method adjusts the parameters of the TF
relation, which in turn modifies the predicted tangential
ellipticities, until they agree {\em in the mean\/} with the observed
tangential ellipticities.  In principle, one could also measure the
scatter in the TF relation by subtracting in quadrature the dispersion
in the ellipticity component measured along the axis oriented
45\arcdeg\ from the line connecting lens and source from the
dispersion in the tangential ellipticity component. In practice,
however, the intrinsic ellipticity dispersion is much larger than the
scatter in predicted shear.  As a result, one cannot easily measure
any source of scatter, be it intrinsic (scatter in the TF relation,
aspherical halos or other random deviations from the assumed
isothermal) or observational (magnitude errors, redshift errors).

The mass model presented produces an unrealistic shear field for the
following reason: because the model assigns mass to galaxies within
the HDF area, and no mass outside the HDF area, there is a change in
the surface mass density across the HDF boundary, the smoothness of
which depends on the scale size $s_{200}$ of galaxy halos.  This leads
to a spurious shear near the HDF boundary.

We correct this bias by adding, for each lens--source pair, 1000
`mock' lenses with the same $s$ and $z_{\rm l}$.  These mock lenses
are placed outside the HDF area but within a radius equal to the
distance from the source to the farthest edge of the HDF area.  The
mock lenses are assigned a $V$ such that the mean surface mass
densities of the HDF lenses and the mock lenses are equal.  The
circular symmetry guarantees that there is no mean shear due to the
mean density field.  The large number of mock lenses guarantees that
the mock lenses do not act as an additional source of noise.

Of course, in reality there will also be density fluctuations outside
the HDF area.  This large-scale structure will generate a
slowly-varying shear across the HDF field.  The addition of the mock
lenses only corrects the surface mass density outside the HDF area to
that within the HDF area, it does not account for possible
fluctuations.  The effect of such large-scale shear is examined in
\secref{lss}.

\section{Monte--Carlo Simulations}
\label{sec:monte}

In order to test for potential biases in the maximum likelihood
method, we have run a series of Monte Carlo simulations.  We choose
galaxies with redshifts from the observed distribution, but assign
random positions within a circular area which is much larger than the
HDF area.  We then adopt a TF relation and assign circular velocities
and truncation radii to the lenses based on their absolute magnitudes.
Galaxies are assigned ellipticities drawn from \eqref{gaussellip}.
Each galaxy's ellipticity parameters are then perturbed by {\em all\/}
foreground galaxies.  Thus our Monte Carlo simulations contain extra
sources of shear which affect the real data, namely lenses outside the
HDF area and lens-source pairs with separations less than $10 \hkpc$
or with redshift separations less than 0.5.  From this supersample,
only the lenses (restricted to $0 < z < 0.85$) and sources within the
HDF area are output. These catalogs are then analyzed as described
above for the real data.  In our simulations we used $V_{\rm f} = 200
\kms$, $s_{200} = 100 \hkpc$ and $\eta = 0.25$.

The Monte Carlo simulations show that the maximum-likelihood method
described above recovers unbiased values of $V_{\rm f}$ ($\la 5\%$)
and $\eta$.  The returned values of $s_{200}$ are also unbiased, but
have a very large scatter, indicating that it is very difficult to
constrain this parameter.  Furthermore, the recovered values of
$V_{\rm f}$ and $s_{200}$ are slightly anti-correlated in the sense
that the halo mass within a radius of $\sim 30 \hkpc$ is the most
tightly constrained combination. If we do not add mock lenses outside
the HDF area, we recover essentially unbiased results for $V_{\rm f}$
but $s_{200}$ is biased low by a factor $\sim 5$.

We can also use the Monte Carlo simulations to test the effect of the
random errors in the photometric redshifts.  We proceed as above,
except that we add errors to the redshifts as follows.  First, we add
a random Gaussian error of 20\% in $z$.  Then we allow for
non-Gaussian tails in the error distribution by assigning a redshift
at random within the range 0 to 5 for 10\% of the galaxies.  We use
the perturbed photometric redshifts to adjust absolute luminosities
accordingly.  We find no significant bias in the recovered value of
$V$ compared with the simulations with no redshift errors.  This
occurs because, for a lens galaxy of fixed apparent magnitude and a
source at a fixed separation from it on the sky, as the redshift
increases, $L$ and hence $V$ increase but there is a compensating
decrease in the ratio $D_{\rm ls}/D_{\rm s}$.  For the default case
$\eta = 0.25$ and for the typical lens and source redshifts studied
here, the two terms very nearly cancel.

However, we do find a small bias in the recovered TF slope: $\eta =
0.22\pm0.01$ compared with the input $\eta = 0.25$.  This bias is much
smaller than the random errors, so we neglect it in this analysis.

\section{Results}
\label{sec:results}

When we fix $s_{200} = 50 \hkpc$, $\eta = 0.35$ (as found for the
local $B$-band TF), and $\zeta = 0$ (i.e.\ no evolution), we obtain
our main result
\begin{equation}
  V_{\rm f}  = 210 \pm 40 \kms
\end{equation}
For this solution, we obtain $\chi^2 = 7.2$, which is significant at
the 99.3\% confidence level ($2.7 \sigma$).  The error bars have been
determined in the usual way, i.e.\ they correspond to $\chi^2 -
\chi^2_{min} = 1$.  Note that the control tests of \secref{control}
confirm that the $\chi^2$ statistic is indeed correct. We then allow
each of the parameters $\eta$, $s_{200}$ and $\zeta$ to be free in
turn.  The best fit values are $\eta = 0.62$, $s_{200} = 12.2 \,
\hkpc$ (with $V_{\rm f} = 250 \kms$) and $\zeta = 1.88$, but in each
case the change in $\chi^2$ is not statistically significant compared
to the increase of 1 degree of freedom.  \figref{likeb} shows the
joint likelihood contours for $V_{\rm f}$ and each of $s_{200}$,
$\eta$ and $\zeta$, with the other two parameters held fixed at their
fiducial values.  The joint likelihood of $\eta$ and $\zeta$ is also
shown.  Note that a flat ($\eta = 0$) TF relation, corresponding to
constant circular velocity independent of luminosity, is not favored
by the data.

\vbox{%
\begin{center}
\leavevmode
\hbox{%
\epsfxsize=8.9cm
\epsffile{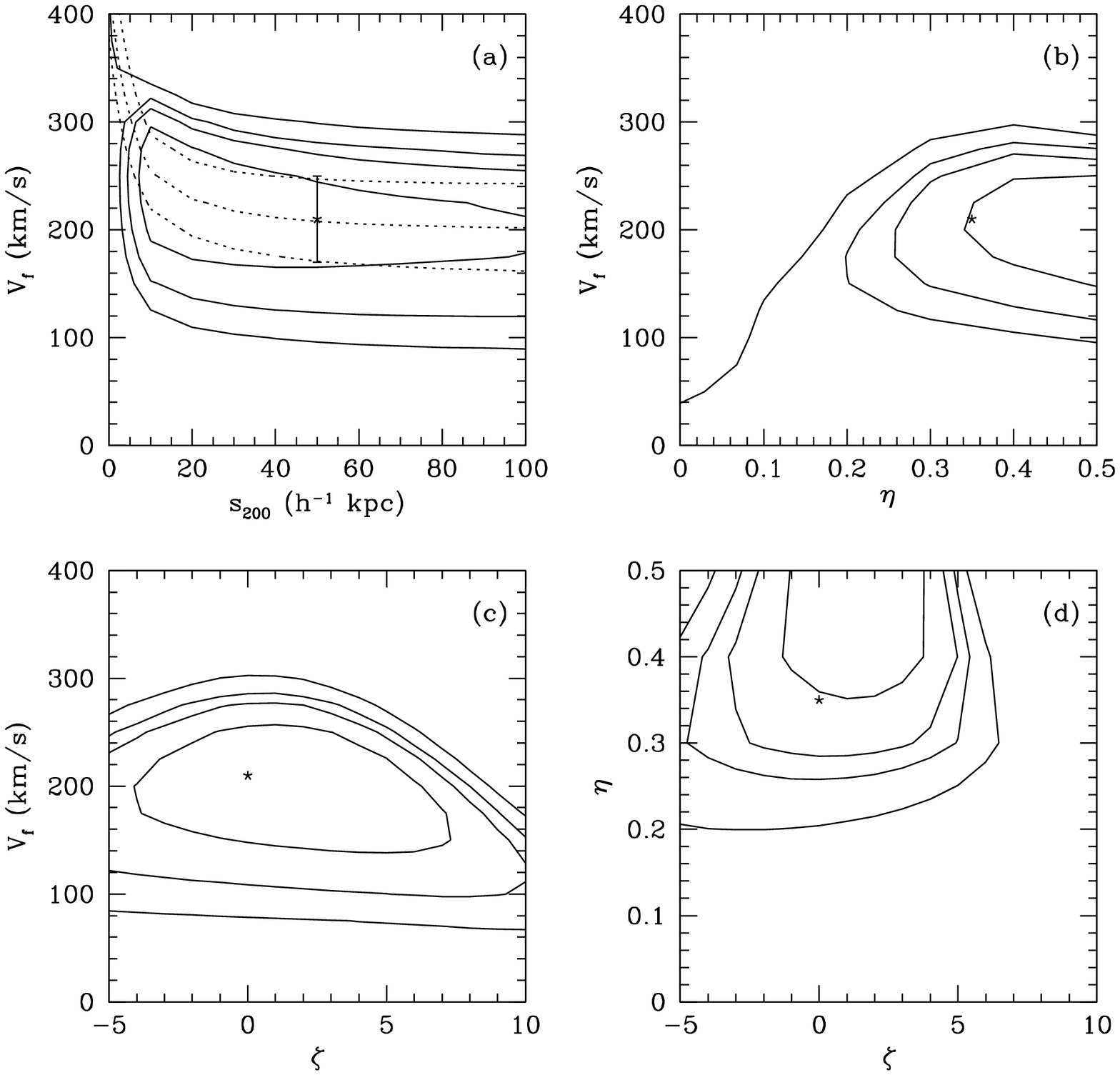}}
\begin{small}
\figcaption{%
  Likelihood contours for various combinations of parameters.
  Contours indicate the 68, 90, 95 and 99\% confidence limits on two
  parameters jointly. (a) The $V_{\rm f}$--$s_{200}$ plane.  The star
  shows the fiducial choice of $s_{200}$ and the best fit value of
  $V_{\rm f}$ and $1 \sigma$ errors.  The three dotted lines show
  models of constant mass within 30 \hkpc, equal to
  $2.8\pm1.2\times10^{11}\,h^{-1}\, M_{\sun}$.  (b) $V_{\rm f}$ and
  $\eta$ (c) $V_{\rm f}$ and $\zeta$ (d) $\eta$ and $\zeta$ with
  $V_{\rm f} = 200 \kms$.
\label{fig:likeb}}
\end{small}
\end{center}}

In \figref{likeb}a, the dotted lines show the loci of truncated
isothermal models with constant mass within $30\,\hkpc$, $M_{30}$.
These lines run close to the isolikelihood lines.  Our solution
corresponds to $M_{30} = 2.8\pm1.2\times10^{11}\,h^{-1}\,M_{\sun}$.
The corresponding $B$-band mass-to-light ratio within this radius is
$75\pm30 \,h\,(M/L_B)_{\sun}$.  However, since we cannot constrain the
total extent of the halos, this number should be taken as a lower
limit to the mass-to-light ratio of galaxies.

For the ellipticity distribution (\ref{eq:gausslorellip}), we obtain
similar results: $V_{\rm f} = 235 \kms$, with $2\,(\log {\cal L}_{\rm
  max} - \log {\cal L}_{V = 0}) = 11.7$.  However, because
\eqref{gausslorellip} is not a Gaussian, this likelihood statistic is
no longer simply related to $\chi^2$, and hence it is difficult to
evaluate the significance of this result and its error.  However, the
tests described in \secref{control} below indicate that the
significance of this result is slightly better than for the default
Gaussian distribution used above.  Consequently, the error bars are
likely to be similar to those quoted above.

\section{Characteristics of the Lens Sample}
\label{sec:new}

In order to interpret our results, it is important to know the
properties of the galaxies which dominate the lensing signal.  The
composition of the lens catalog itself does not accurately reflect the
galaxies which dominate the lensing signal because several factors,
most importantly luminosity and redshift, determine whether a galaxy
is an efficient lens. In general, the lens-source pairs with the
highest predicted shear, $|\gamma_{t,ij}|$, should have the greatest
effect on the likelihood.  We can therefore determine the `effective'
mean, $\tilde{x}$, of any quantity $x$ by weighting the lens-source
pair $w_{ij} = |\gamma_{t,ij}|$:
\begin{equation}
  \tilde{x} = \frac{\sum_{ij} w_{ij}\,x} {\sum_{ij} w_{ij}} \,.
  \label{eq:effective}
\end{equation}
We will use this shear-weighting method to determine the radial
dependence of the lensing signal, and to obtain the typical redshifts,
magnitudes and colors of the lens galaxies.

We can get an impression of the radial dependence of the galaxy-galaxy
lensing signal by calculating the effective mean tangential
ellipticity of the sources in all lens-source pairs as a function of
separation in $\hkpc$.  The ellipticities are weighted according to
the predicted shear, as described above.  This is plotted in
\figref{avshear} with the predictions corresponding to combinations of
$V_{\rm f}$ and $s_{200}$ in the maximum likelihood valley of
\figref{likeb}a. The small difference in the shapes of these curves
illustrates the difficulty of constraining $s$.

\vbox{%
\begin{center}
\leavevmode
\hbox{%
\epsfxsize=8.9cm
\epsffile{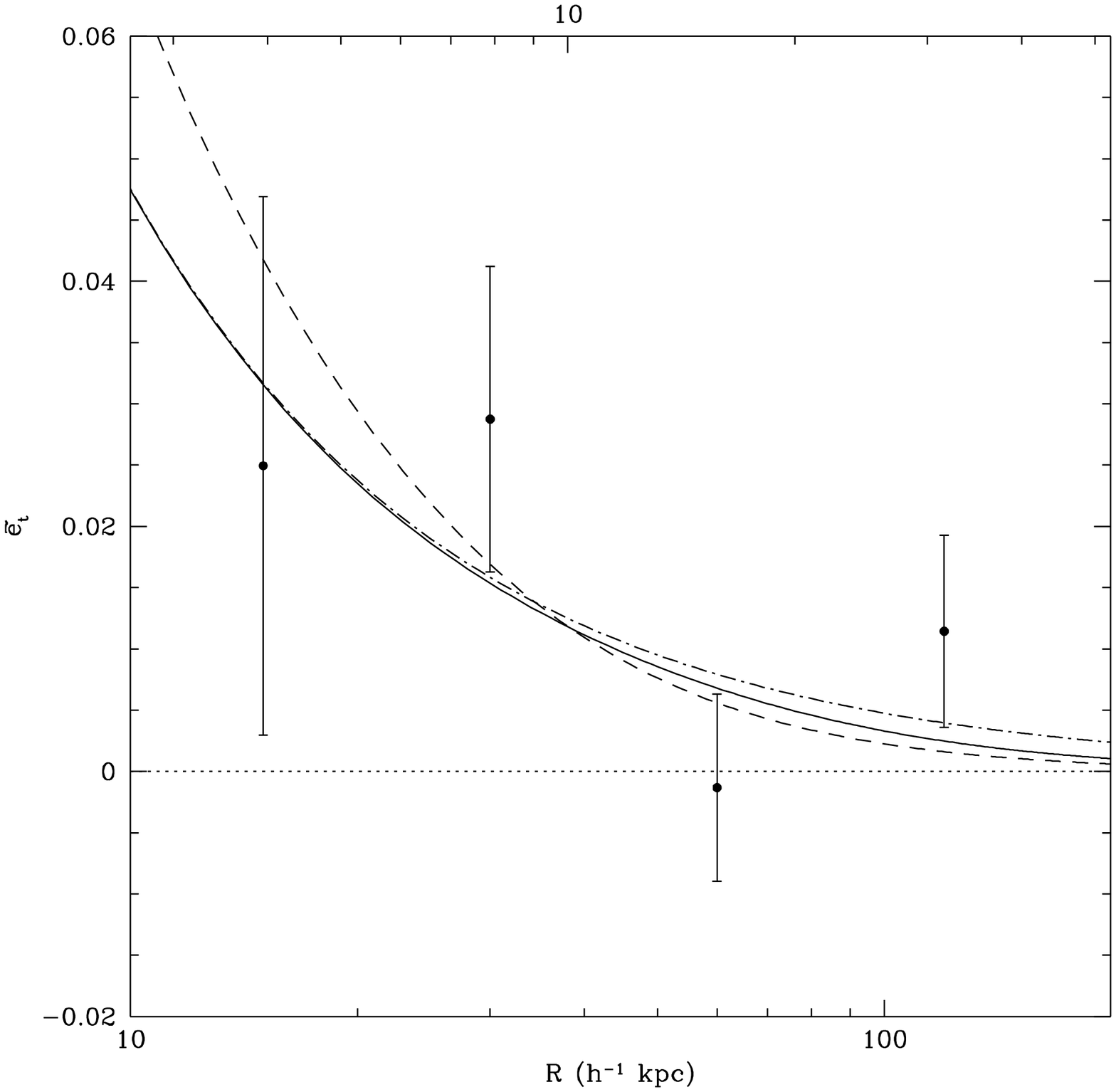}}
\begin{small}
\figcaption{%
  The weighted mean tangential ellipticity of sources around all
  lenses as a function of separation in \hkpc.  The top axis gives the
  separation in arcseconds for $z_{\rm l} = 0.6$.  The weight for each
  pair is proportional to the predicted tangential shear; see text for
  details.  The solid curve shows the change in ellipticity of a
  source at $z = 2.4$ due to gravitational shear of a lens at $z =
  0.6$ for the fiducial model $V_{\rm f} = 210
  \kms$, $s_{200} = 50 \hkpc$. The dashed curve shows a model with
  $V_{\rm f} = 250 \kms$, $s_{200} = 12.5 \hkpc$ and the dot-dash line
  shows a singular isothermal sphere with $V = 210 \kms$ and no
  truncation radius.
\label{fig:avshear}}
\end{small}
\end{center}}

We can also use shear-weighting to determine the characteristics of
lens and source galaxies. The effective mean redshifts are
$\tilde{z_{\rm l}} = 0.6$ and $\tilde{z_{\rm s}} = 2.4$.  The weighted
lens and source redshift histograms are shown in the lower two panels
of \figref{zhist}.  The effective mean lens magnitude is $\tilde{M_B}
= -18.5$. The lensing signal arises equally from 2 subsamples: 14
bright lenses with $M_{B} \la -18.5$ (the brightest of which has
$M_{B} = -20.3$) and the remaining 194 faint lenses with $M_{B} \ga
-18.5$ (mean $-17.5$, but with a tail extending to $-15$). It is worth
noting that no single lens galaxy is responsible for the entire signal
detected here.  The lens galaxy with the most impact on the solution
is the $I_{\rm ST} = 20.87$ mag, $z = 0.518$ galaxy at $x=1396,
y=1167$ on WF3.  Our result remains significant, however, if this
galaxy is removed.
The shear-weighted shear for a typical lens-source pair is
$\tilde{\gamma_{ij}} \sim 0.02$, and hence the sample contains the
equivalent of $\tilde{N_{\rm ls}} \sim 1000$ of such pairs.  In a
similar fashion, we infer that the effective number of lenses is
  $\tilde{N_{\rm l}} \sim 40$.
The effective mean pre-seeing shear polarizability is
$\tilde{P^{\gamma}} = 0.74$.

We have also investigated the morphological types and colors of the
lens galaxies.  Morphological classifications are available only for
the brightest ($I < 25$) HDF galaxies (van den Bergh et al.\ 1996).
We have identified all galaxies classified as E or S0 by both van den
Bergh and Ellis.  We have compared results obtained from the $I < 25$
subsample of our lens sample (73 galaxies) with and without the 11
early-type galaxies.  We find only a 3\% drop in $V_{\rm f}$ when the
early-type galaxies are excluded, although the significance
of the result drops slightly, to the $2.5\sigma$ level.

We have also measured rest-frame \ub\ colors of the HDF lens galaxies
and compared these with the colors of early-type galaxies ($T < 0$)
from RC3.  Approximately 90\% of early-type galaxies in RC3 have $\ub
> 0.23$.  If the stars in ellipticals were formed at high redshift,
then we expect passive evolution to shift this limit to $\ub > 0.12$
at $z = 0.6$ (Worthey, 1994).
The median rest-frame \ub\ color of the 11 morphologically classified
early-type galaxies is 0.15. In the HDF lens sample, only $11\%$ of
galaxies have $\ub > 0.12$.  This fraction is similar to the fraction
(13\%) of galaxies classified morphologically as E or S0.  If these
red galaxies are excluded from the lens catalog, $V_{\rm f}$ is
unchanged.  Thus selection by morphology and selection by color both
lead to the conclusion that our result is not affected by presence of
a small fraction of early-type galaxies.  In general, there is little
evidence of color dependence. The red half of the sample has a higher
$V$ than the blue half by only $16\pm32$ per cent.  Larger samples are
required to test the color dependence in more detail.

In summary, we conclude that the lensing signal is dominated by spiral
galaxies at $z \sim 0.6$, with luminosities 1 mag.\ below $L_*$.  Thus
our lens sample is similar to the $I$-band limited CFRS redshift
survey (Lilly et al.\ 1995).

\section{Possible Systematic Effects}
\label{sec:sys}

This section describes tests performed to explore possible systematic
effects which might bias our results.

\subsection{Control Tests}
\label{sec:control}

We have performed a number of control tests in order to verify the
reality of the signal.  If we rotate the source position angles by
45$^{\circ}$ we expect no signal, and indeed we find none.  We have
also experimented with randomizing the position angles of the sources,
and with randomizing the positions of the sources and lenses.  For
each of these tests, we generate 1000 such random realizations.  We
obtain a signal as significant as the observed signal in 18, 17 and 2
cases, respectively. These tests confirm that our result is
significant at the $99\%$ confidence level, in agreement with the
significance level obtained directly from the $\chi^2$ distribution.

If we adopt the modified Gaussian ellipticity distribution
\brref{gausslorellip}, we find that the number of false positives is
reduced by a factor of 2, leading to an improved significance level of
$99.5\%$.

We have also examined the subsample of 925 sources with $I_{\rm ST} >
28$.  These galaxies are too faint to have reliable photometric
redshifts.  However, they can still be used, albeit with loss of
precision and some systematic uncertainties.  We have performed Monte
Carlo simulations in which the faint galaxies are assigned redshifts
drawn at random from the redshift distribution of bright sources.  For
any one realization, we do not necessarily obtain a significant
detection, but when the $\chi^2$ values are averaged over all
realizations we find a minimum with $\chi^2 = 3.5$ at $V_{\rm f} =
165^{+35}_{-45}$ \kms.  Thus, from the faint sources alone, we detect
a marginally significant signal (at the $94\%$ confidence level) and a
value of $V_{\rm f}$ which is consistent with that found from the
brighter sources.  However, we prefer not to place too much weight on
the derived value of $V_{\rm f}$ since it depends on the uncertain
redshift distribution of these very faint source
galaxies%
\footnote{Note that if the faint sources are in fact at greater
  distances than the bright sources, as is likely, then by this
  procedure we have underestimated $D_{\rm ls}/D{\rm s}$ and hence
  have overestimated $V_{\rm f}$. However, the correction factor is
  likely to be small: if, for example, the faint sources lie at a
  typical $z$ of 3 (versus 2.4 for the bright sources), then $V_{\rm f}$ is
  overestimated by only 3 per cent.}.  The significance of the
detection should be less sensitive to the shape of redshift
distribution, so we take this result as independent confirmation of
our detection using the brighter sources.  Taking both bright and
faint results together, the significance of the detection is 99.8\%.
It would appear that photometric redshifts increase the statistical
weight of sources by a factor $\sim 3$ relative to no redshift
information.

\subsection{Effects of PSF Anisotropies}
\label{sec:psf}
It is likely that the drizzling algorithm does not completely remove
the anisotropy of the HST PSF.  We have examined the mean ellipticity
of source galaxies measured tangentially to the chip center, as a
function of radius from the center.  We find a weak, marginally
significant ($2 \sigma$), and approximately linear trend with radius.
The mean tangential ellipticity increases from zero at the center to
$0.018\pm0.008$ at the edge.  (Note that we find no mean ellipticity
tangential to the {\em field\/} center.)  This anisotropy is in the
same sense as that found by Hoekstra et al.\ (1997), although the
amplitude is apparently somewhat smaller for than they found for
undrizzled HST images.  In order to test how this might affect our
results, we have constructed a modified source catalog with the mean
trend subtracted from the measured ellipticities.  The difference in
mean tangential ellipticity between the modified catalog and the
original catalog is negligible except for pairs with separations $\ga
30$ arcseconds ($\ga$ half the WF chip size).  Consequently, we find
that this modification has no effect on the value of $V_{\rm f}$.
However, for the modified ellipticity sample, the very large
separation ($> 30$ arcseconds, roughly equivalent to $> 100 \hkpc$)
pairs tend to have negative mean tangential ellipticities.  If these
pairs are used in the analysis, the likelihood derived from the
modified source catalog marginally favors mass models with small
truncation radii, $s_{200} \sim 5\,\hkpc$ with a corresponding
increase in $V_{\rm f}$ (see, for example, the dashed line in
\figref{avshear}).  Such models are preferred because the predicted
shear goes to zero more quickly at large radii.  This preference for
small truncation radii does not occur with the original ellipticities,
or for either sample if pairs are limited to a maximum of 30
arcseconds separation. Ultimately, we do not know whether the trend
found above is an uncorrected anisotropy or not.  We prefer avoid this
issue by limiting the sample used in this paper to the robust pairs
with separations $< 30$ arcseconds.

\subsection{Isophotal contamination}
One possible concern is that the extended luminosity profile of bright
galaxies might affect the measured shapes of the faint galaxies in
their wings.  We have explored this possibility by adding artificial
faint galaxies in the wings of bright ($I_{\rm ST} < 25$) HDF
galaxies, using the IRAF artdata/mkobjects package.  The artificial
galaxies have a range of sizes, surface brightnesses and ellipticities
consistent with those in our source catalog.  
The simulated source galaxies are placed with random position angles
and at various radii from the centroids of the bright HDF galaxies.
We then extract their image parameters using imcat and compare these
to the input parameters.

\vbox{%
\begin{center}
\leavevmode
\hbox{%
\epsfxsize=8.9cm
\epsffile{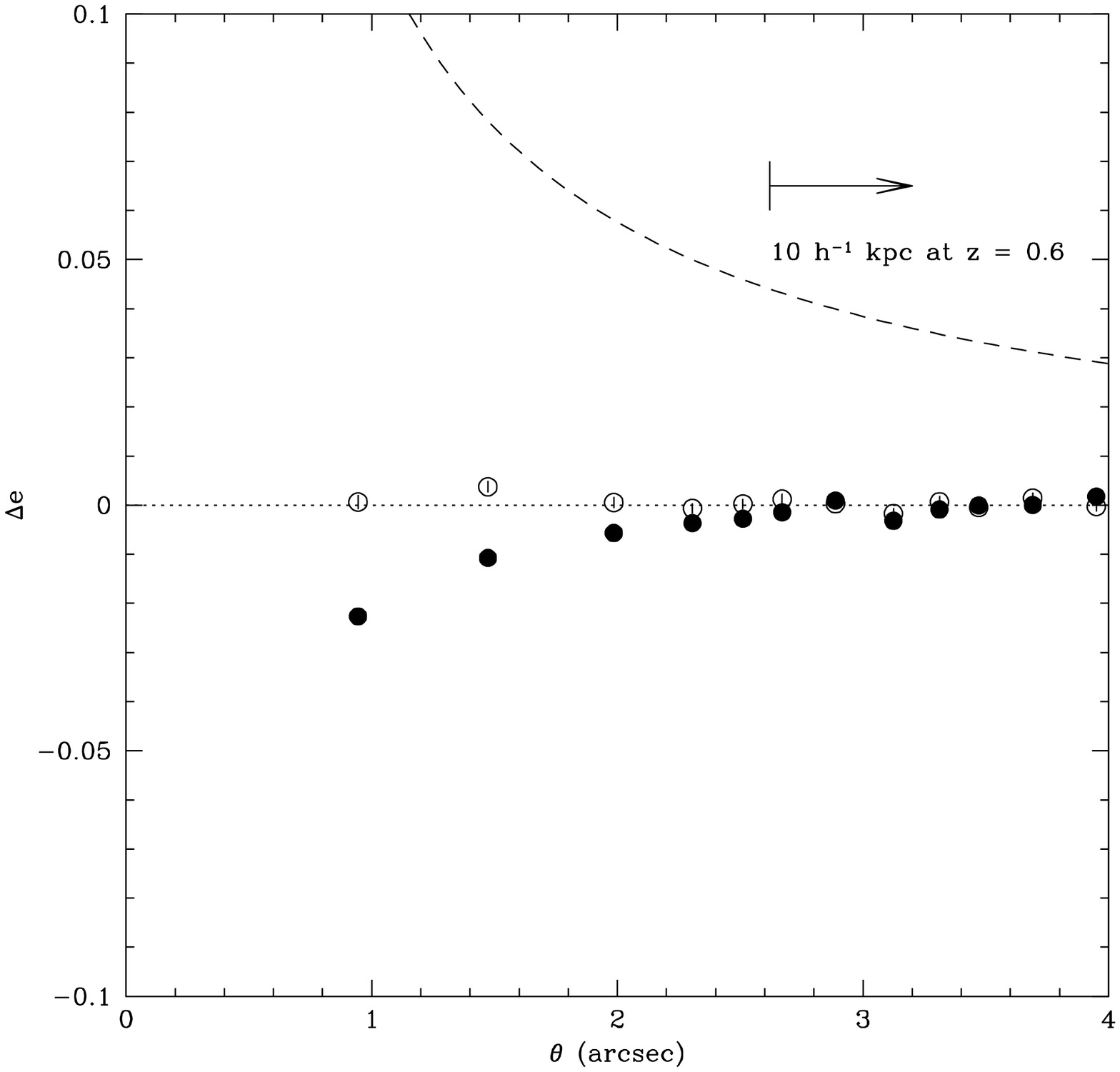}}
\begin{small}
\figcaption{%
  A simulation of the effect of light
  contamination from foreground galaxies on the shape parameters of
  background galaxies.  The change in imcat ellipticity of
  simulated source galaxies is plotted as a function of separation
  from bright galaxies ($I_{\rm ST} < 25$).  Solid points indicate the
  change in ellipticity along the direction tangential to the line
  joining the pair.  Open points show the change in the direction
  rotated by 45$^{\circ}$.  These are consistent with zero as
  expected.  The dashed line shows the expected change in ellipticity
  due to weak lensing from a lens with $V = 200$ \kms\ at $z = 0.6$
  on a background source at $z = 2.4$.  Note that the contamination
  effect is small and goes in the opposite sense to the lensing
  signal: galaxies tend to be biased towards radial alignments.
\label{fig:contam}}
\end{small}
\end{center}}

\figref{contam} shows the median change in imcat ellipticity as a
function of projected separation in arcseconds. The solid points show
the change ellipticity along the tangential direction, and the open
symbols show the change along a direction rotated by 45$^{\circ}$.
For comparison, the dashed line shows the change in tangential
ellipticity due to a $V = 200 \kms$ isothermal sphere at $z_{\rm l} =
0.6$ for source at $z_{\rm s} = 2.4$.  From \figref{contam} it can be
seen that the effect is of opposite sign to the effect of
gravitational lensing: isophotal contamination distorts the faint
objects in the radial direction.  A very similar effect is found for
ellipticities measured with SExtractor. The effect is many times
weaker than the effect of gravitational lensing at all radii and is
completely negligible for the minimum separation of $10\hkpc$ ($\sim
2.5$ arcseconds) used here. We are therefore justified in neglecting
this effect.

\subsection{Satellites and Spiral Arms}
Another potential systematic effect is the contamination of the source
catalog with satellite galaxies or spiral arm segments which have
been treated as separate objects by the object finding software.  It
is not known whether satellites tend to be preferentially tangentially
oriented with respect to the primary, but spiral arm segments
certainly would be.  If such objects are included in the source
catalog, they could bias the results.

We can use cuts in projected separations to test for such
contamination.  The default minimum projected metric separation is $10
\hkpc$.  However, as a check, we have experimented with minimum
separations in units of the major-axis second moment.  When the
minimum separation is set to 5 times the major-axis second moment
(equivalent to $\sim 8.7$ exponential disk scale-lengths), we find an
equally strong lensing signal.  Since this separation is well beyond
the optical radius, there is little chance that spiral arms are
contaminating our signal.

A great advantage of this study arises from the use of photometric
redshift data, which allows us to cleanly separate foreground and
background objects.  Note that we impose a minimum redshift separation
between lens and source of $0.5$, which corresponds to $\sim 3$ times
the photometric redshift error at the effective lens redshift of 0.5.
It is worth noting that if the minimum redshift separation is
decreased to 0, the significance of our result increases to $3\sigma$
and $V_{\rm f}$ increases by 5\%. However, if the minimum redshift
separation is increased to 1, we still find a significant signal.
These tests confirm that our results are not biased by objects at the
same redshift as the lens.

\subsection{Effects of Large-Scale Structure} 
\label{sec:lss}
It is well known that a uniform sheet of mass produces no shear, and
hence cannot bias our results.  In \secref{method}, however, we noted
that density fluctuations near the HDF would give rise to a
slowly-varying shear across the HDF field. If the lens-source pairs
sample all position angles uniformly, it is easy to show that a
uniform shear field, or one with a constant gradient, will drop out.
Thus we expect that the slowly-varying signal expected from
large-scale structure will not affect our results.

By taking the mean of the ellipticities on all three WF chips, we find
a marginally significant tendency for galaxies to be elongated in the
$x$-direction (i.e. along the line connecting the bases of chips WF3
and WF4) with mean ellipticity $0.014\pm0.007$.  If this were due to
large-scale shear, the corresponding value of $\gamma$ would be
$0.02\pm0.01$.  This is in good agreement with the r.m.s.\ mean shear
expected in the non-linear regime (i.e. on scales of 1 arcmin) from
sources at $z = 2$ ($\sim 0.03$, Jain \& Seljak 1997).  Alternatively,
it may arise from incorrect registration during the drizzling
procedure.  In either case, it is interesting to know whether this may
bias our results.  
If we subtract off the mean ellipticity, we find that $V_{\rm f}$ is
unchanged.  Note that for pairs with larger separations, the
lens--source position angles are no longer random.  The largest
separations are dominated by pairs between WF2 and WF4, which have
lens--source position angles $\pm45^{\circ}$.  Nevertheless, even if
we include these large separation pairs, our $V_{\rm f}$ result is
essentially unaffected.

\subsection{Effect of correlated dark matter on intermediate scales}

It is also possible our results might be biased by mass concentrations
other than the halos of the observed HDF galaxies. It is easy to see
that dark matter concentrations which are uncorrelated with the
observed galaxies will not affect our results: the shear produced by
such concentrations will add scatter to the source ellipticities ---
something which is difficult to separate from the intrinsic scatter in
ellipticity --- but will add no net tangential shear.  On the other
hand, dark matter which is correlated with bright galaxies could
potentially bias our results.  We consider here two possible sources:
undetected galaxies which are clustered with the observed galaxies in
the lens sample, and dark matter on group scales.

From the spatial two-point correlation function for galaxies, we can
calculate the mean excess number of $\sim L_*$ neighbors within a
given distance of a galaxy.  The distance where that mean excess is
unity is approximately 300 \hkpc\ at the present epoch, and at smaller
distances the probability of having a correlated neighbor is small.
At high redshift, one has to allow for evolution: the number density
goes up, but the correlation amplitude goes down, roughly
compensating.  For the $\sim 30$ \hkpc\ scale which dominates our
signal, the probability of finding a neighbor remains small.
Furthermore, note that at $z \sim 0.6$, all galaxies with $M_B < -15$
are included in the lens sample and hence in the mass model. Thus the
contamination from galaxies fainter than this limit will be smaller
still.

Another candidate for correlated dark matter is the dark matter
distributed in the groups and clusters in which (at least some of)
these galaxies dwell. We have examined the effect of group-scale dark
matter concentrations on our results.  Specifically, we have
calculated the mean tangential shear around a galaxy, as a function of
galactocentric radius $R$, due to a spherically symmetric group
located at a distance $R_{\rm gr}$ away from the galaxy.  The group is
assumed to have a truncated isothermal mass density profile.  Clearly,
if $R_{\rm gr} = 0$ then the circular velocity of the group dominates
the contribution from the galaxy.  On the other hand, for large values
of $R_{\rm gr}$, we find that the group typically has a negative
effect on the mean tangential shear over the range of galactocentric
radii considered here.  This is illustrated graphically in
\figref{group}.  The overall net effect depends on the distribution of
galaxies with respect to the group center, i.e.\ on the distribution
of $R_{\rm gr}$.  If galaxies are unclustered with respect to the
group center, then there is no net effect, for the reasons given
above. If on the other hand, galaxies are distributed like the group
dark matter, that is as $r^{-2}$, or $R^{-1}$ in projection, then the
group contributes a small net positive contribution to tangential
shear, the amplitude of which is independent of $R$ and which depends
on the circular velocity and cut-off radius of the group.  We find
that, for our maximum likelihood method and default parameters, the
bias to the galaxy circular velocity is $\sim 0.02 \, (V_{\rm gr}/V)^2
\,[s_{\rm gr}/(300 \hkpc)]^{-1}$, with a weak dependence on the
assumed $s$.

\vbox{%
\begin{center}
\leavevmode
\hbox{%
\epsfxsize=8.9cm
\epsffile{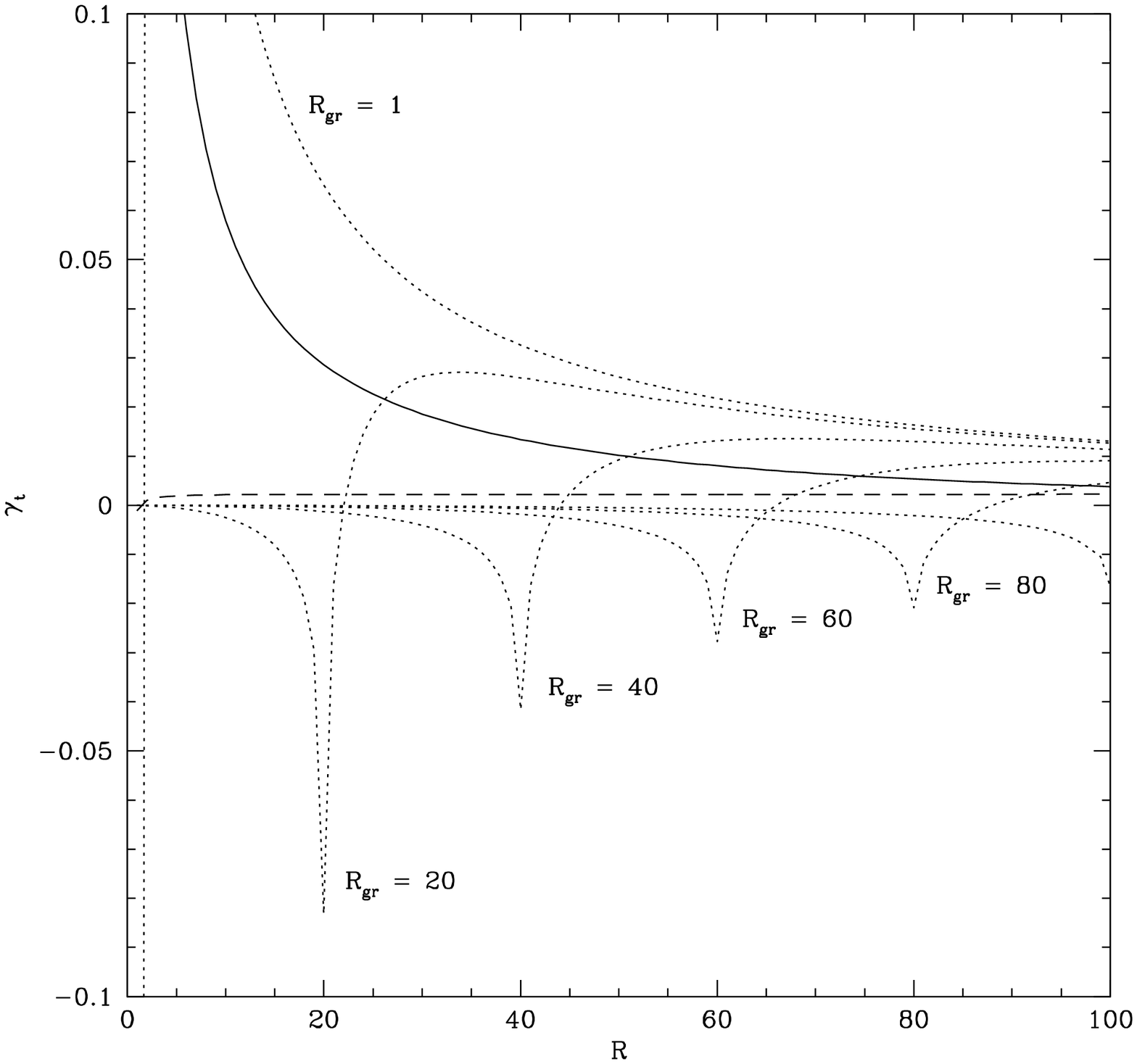}}
\begin{small}
\figcaption{%
  The tangential ellipticity with respect to a galaxy as a function of
  galactocentric radius $R$ due to a $V = 200$ \kms\ galaxy
  (solid line) and the contributions from a group with $V_{\rm gr} =
  300$ \kms\ and truncation radius $s_{\rm gr} = 300$ \hkpc, where the
  galaxy lies at
  separations $R_{\rm gr} = 1, 20, 40, 60, 80, 100$ \hkpc\ from the
  center of the group (dotted
  lines).  The dashed line shows the mean contribution from the group
  averaged over all galaxies in the group, if the galaxies are
  distributed like the mass in the group.
\label{fig:group}}
\end{small}
\end{center}}

To apply this formula, we need the typical circular velocity
dispersion of groups as well as their typical sizes.  While the {\em
  pair-weighted\/} r.m.s.\ velocity dispersion of galaxies is
$540\pm180$ \kms (Marzke et al.\ 1995), this value is not
representative of the field because it is biased by the large number
of pairs in rich clusters.  A single-galaxy weighted estimate of the
line-of-sight velocity dispersion of field galaxies with respect to
their local 2 \hmpc\ neighborhood is $\sim 100$ \kms\ (Davis, Miller
\& White 1997).  If all field galaxies are assumed to lie in groups
then the corresponding circular velocity of the group would be $\sim
150$ \kms.  A more realistic assumption might be to assign 50\% of
galaxies to groups with typical circular velocities of $\sim 250$
\kms.  In either case, the mean bias is identical: it is negligibly
small.

\section{Discussion}
\label{sec:discuss}

\subsection{Comparison with previous galaxy--galaxy lensing results}
Our results are in good agreement with the original detection of BBS,
who found $V_{\rm f} = 220 \pm 80 \kms$ for a 
fiducial $M_V = -20.3$ galaxy%
\footnote{The fiducial magnitude $M_r = -18.5$ quoted in BBS is
  incorrect (T. Brainerd, private communication).} (with TF slope
$\eta = 0.25$).  If we adopt a mean rest-frame $B-V$ color of 0.6 for
galaxies at $z \sim 0.4$, then the BBS fiducial magnitude corresponds
to $M_B = -19.7$.  Scaling their $V_{\rm f}$ to our fiducial magnitude
($M_B = -18.5$) with $\eta = 0.25$ then yields $165\pm60 \kms$.  This
is consistent with our result.  Note that BBS make no correction for
the circularization of their sources due to seeing. As a result they
will tend to underestimate the shear and hence the circular velocities
of the lens galaxies.

At face value, our results also agree well that of DT who found
$\expec{V} = 260\pm45 \kms$. However, our analysis differs
significantly from theirs in a number of ways.  We use photometric
redshifts and scale $V$ according to the TF relation, whereas DT
adopted a constant $V$ for all bright ($22 < I < 25$) galaxies, and
assumed a redshift distribution for lenses and sources.  If, following
DT, we also assume a flat ($\eta = 0$) TF relation, we do not detect a
significant signal.  This may be a result of the different range of
projected separations probed: DT use separations less than 5
arcseconds, which corresponds to 16 \hkpc\ at their mean lens redshift
of 0.4, whereas we specifically limit our sample to separations larger
than 10 \hkpc.

\subsection{Comparison with the Tully-Fisher relation}
                       
In \secref{results}, we found that our sample was not capable of
measuring the $B$-band evolution of the TF relation based on the HDF
data alone.  In \secref{new}, we showed that our signal was dominated
by spiral galaxies with luminosities 1 mag.\ below $L_*$.  It is
therefore interesting to compare our result to the Tully-Fisher
relation derived from HI linewidths nearby and optical rotation curves
at intermediate redshifts.  However, in order for this comparison to
be valid, we must first address two technical issues (corrections to
total magnitudes and inclination corrections) and the more general
issue of the flatness of the rotation curves at large radii.

The magnitudes in our lens catalog are isophotal, whereas total
magnitudes are used in Tully--Fisher studies. At the typical lens
galaxy redshift of 0.6, the limiting $I_{\rm ST}$ isophote is close to
a rest-frame isophote of $25\,B\, {\rm mag}/{\rm square\, arcsecond}$.
At low redshift, the correction from this isophotal magnitude to total
$B$ magnitude is only 0.11 mag.\ for spiral galaxies (de Vaucouleurs
et al.\ 1991, hereafter RC3).  Note that the correction from isophotal
magnitude to total magnitude depends only on the central surface
brightness of the galaxy and the surface brightness of the isophotal
limit. For an Sbc galaxy with a bulge-to-total luminosity fraction of
0.25 and a central disk surface brightness of $21.65 \,B\, {\rm
  mag}/{\rm square\, arcsecond}$, the correction is $\sim 0.15$ mag,
in reasonable agreement with the RC3 value.  However, if intermediate
redshift galaxies have higher central surface brightnesses by 0.5--1.0
mag., as suggested by some studies discussed below, then this
correction falls to 0.10--0.07 mag. We adopt 0.1 mag.\ for the
isophotal-to-total magnitude correction.

Our lens catalog contains galaxies of all inclinations.  Rather than
correct each HDF galaxy individually, we correct the fiducial
magnitude using the prescription of Tully \& Fouqu\'{e} (1985) by
averaging over all inclinations.  Thus, with both corrections the
fiducial face-on total magnitude is $M_B = -18.9$.

Finally, note that the lensing method probes the mass within $\sim 30
\hkpc$, whereas studies of optical rotation curves or HI linewidths
probe out to the edge of the optical or HI disks ($\sim 10-20 \hkpc$).
If rotation curves are not flat then an offset might be expected
between the two measures.  The resolved HI rotation curves of Broeils
(1992) extend out to $30 \hkpc$ for galaxies with $V > 150\,\kms$ and
are flat or gently declining beyond the optical radius (see Fig. 16 of
Courteau 1997).  If the H$\alpha$ rotation curves of Courteau (1997)
are extrapolated to $30 \hkpc$, then for galaxies with $V \sim
200\,\kms$, $V_{\rm max}$ is on average $\sim 5$\% lower than $V(r=30
\hkpc)$.  On average, we therefore expect the lensing circular
velocities to be a reasonable measure to compare with HI linewidths or
optical rotation curves.

For the local TF relation we use the Ursa Major $B$-band data from
Pierce \& Tully (1988).  We assume a Hubble velocity of 1324 \kms\ for
the Ursa Major cluster.  For comparison, we also show the scaling
relation followed by elliptical galaxies.  We have used the $B$-band
Quality `1' data of Faber et al.\ (1989), and multiplied the central
velocity dispersions, $\sigma_{\rm c}$ by $\sqrt{2}$ to obtain
effective circular velocities.  The results are plotted in
\figref{tfb}.  The HDF TF relation is shown by the hatched area.

\vbox{%
\begin{center}
\leavevmode
\hbox{%
\epsfxsize=8.9cm
\epsffile{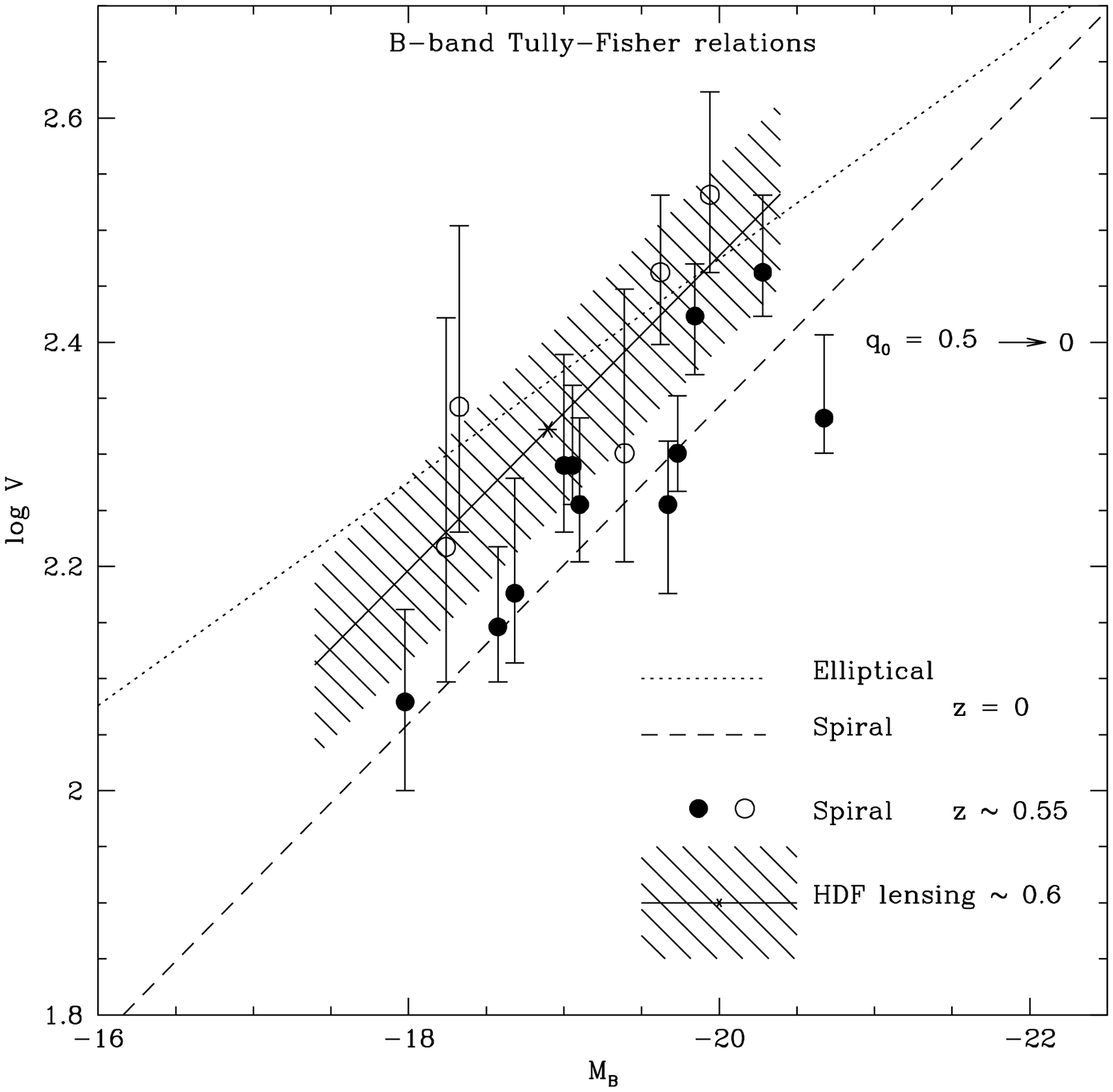}}
\begin{small}
\figcaption{%
  The $B$-band Tully--Fisher relations for nearby spirals (dashed
  line) and ellipticals with $V = \sqrt{2} \sigma_c$ (dotted line).
  The $\expec{z} = 0.54$ data of Vogt et al.\ (1996, 1997) are shown
  by the large circles with error bars (filled symbols for high
  quality data). The mean TF relation for the HDF lens galaxies at $z
  \sim 0.6$ is indicated by the hatched region.  The horizontal
  extent shows the magnitude range covered by the data, and the
  vertical extent gives the $1\sigma$ error in $\log V$.  The small
  arrow shows the effect on our HDF lensing solution if $q_0$ is
  changed from $0.5$ to $0$.
\label{fig:tfb}}
\end{small}
\end{center}}

By comparing the rotation curve and lensing TF relations, we find that
the HDF galaxies are fainter at a fixed $V$ than the local TF
galaxies.  We thus infer $\Delta M_B = +1.0\pm0.6$ from $z = 0$ to $z
\sim 0.6$.  Equivalently, at a fixed luminosity galaxies rotate 38\%
faster at intermediate redshift.  A brightening of more than $-0.2$
mag.\ is excluded at the 95\% confidence level by our data.

If we adopt $q_0 = 0$, $V_{\rm f}$ drops to $190\pm35\kms$.
Equivalently, galaxies brighten by 0.3 mag.\ at a fixed circular
velocity compared to the $q_0 = 0.5$ case, but remain fainter than $z
= 0$ spirals by $0.7\pm0.55$ mag.

For comparison, we also show in \figref{tfb} the intermediate redshift
results of Vogt et al.\ (1996, 1997), converted to $q_0 = 0.5$, $H_0 =
100$ (large circles).  These galaxies have $\expec{z} = 0.54$, close
to our lens sample.  Note that with this choice of cosmology, the
evolution of the Vogt et al.\ sample with respect to the local TF
relation is a dimming $\Delta M_B = +0.21\pm0.16$ (high quality data
only)
or $+0.33\pm0.20$ (weighted fit to all data)%
\footnote{ The difference between these results and the evolution
  $-0.4$ mag.\ quoted by Vogt et al. (1997) for their high-quality
  data arise from the different $q_0$ (they adopted 0.05) and an error
  in the inclination corrections (see Vogt et al. 1998) used by Vogt
  et al. (1996, 1997).}.  Our results are therefore consistent with
the Vogt et al.\ data.  Both results favor little or no luminosity
evolution in the $B$-band for galaxies $\sim 1$ mag.\ below $L_*$.

This conclusion differs from those of Rix et al.\ (1997) and Simard \&
Pritchet (1998). Both of these groups found that the galaxies in their
samples (at $z \sim 0.3$) were brighter than their local counterparts
by $\sim -1.5$ mag.  However, both of these samples are dominated by
galaxies with large \ion{O}{2} equivalent widths: 15 of 19 in Rix et
al.\ and 11 of 12 of galaxies in Simard \& Pritchet have equivalent
widths larger than 20 \AA, indicating high rates of star formation.
Clearly more local TF data, spanning a range of morphologies and star
formation rates, are needed in order to understand the differences in
the TF evolution of the different samples.

The total luminosity {\em density\/} is evolving strongly, $\zeta =
2.7\pm0.5$ in the $B$-band (Lilly et al.\ 1996). The luminosity
density evolution is dominated by galaxies near $L_*$. Because these
same galaxies dominate our lens sample, we can compare the global
luminosity evolution with our result for the evolution of the
Tully-Fisher relation. We can rule out (at the $3.6 \sigma$ level) the
possibility that the evolution found by Lilly et al.\ is solely due to
the luminosity evolution of individual galaxies.  This suggests that
the evolution in the global luminosity density is not dominated by
luminosity evolution of individual galaxies but rather by the
evolution in their number density.

\subsection{Evolution in size or luminosity?}
\label{sec:size}

It is interesting to compare our results with the evolution of surface
brightness, which is a complementary probe of galaxy evolution.

Schade et al.\ (1996a) found that cluster and field ellipticals at $z
\sim 0.6$ had surface brightnesses which were $-0.8$ mag.\ brighter
than $z=0$ ellipticals in the rest frame $B$-band.  Note that if this
is attributed to passive luminosity evolution then in \figref{tfb},
the $z = 0.6$ Faber-Jackson relation would lie on top of the $z = 0$
Tully-Fisher relation. Thus passive evolution might explain the small
difference in $V_{\rm f}$ that we find between our the spiral and
elliptical subsamples.

For spiral galaxies, the situation is more complicated. Schade et al.
(1996b,c) find that spirals at $z \sim 0.6$ have surface brightness
which are $\sim -1.3$ mag.\ brighter than at $z=0$.  Forbes et al.\ 
(1996) find $-0.6$ mag.\ for their sample at $z \sim 0.5$, while Vogt et
al.\ find a similar increase in {\em surface brightness} for their TF
sample.  If this surface brightness evolution is attributed to
luminosity evolution, then the luminosity evolution derived from
lensing disagrees with the results of Schade et al.\ by $\sim 2.2$
mag, and disagrees by $\sim 1.6$ mag.\ with those of Forbes et al. and
Vogt et al.

An alternative interpretation of this surface brightness evolution is
that the population of disk galaxies is undergoing evolution in their
characteristic sizes.  Such a trend is predicted in hierarchical
models if the halos of disk galaxies were assembled recently (say
between $z \sim 1$ and $z \sim 0$). The density at which halos
virialize is a multiple of the critical density at the time at which
they collapsed.  Thus, at a fixed mass, halos which collapse at a
higher redshift will have higher densities than their lower redshift
counterparts.  If the dark matter halo determines the amplitude of the
rotation curve, then the high redshift disks will have higher circular
velocities at a given mass.  Using the prescriptions of Mo, Mao \&
White (1997, see also Mao, Mo \& White 1997), and assuming that the
halo is assembled at the redshift at which it is observed and that the
mass-to-light ratio of the disk is constant, then in an $\Omega = 1$
universe, disks of a fixed circular velocity which form at $z \sim
0.6$ will have scale lengths a factor of 2 smaller, central surface
brightnesses which are 0.75 mag brighter, and total luminosities which
are 0.75 mag.\ fainter compared to disks forming at $z \sim 0$.  These
predictions are similar to our HDF TF evolution, the TF evolution of
Vogt et al.\ and the surface brightness evolution of Forbes et al.\ 
and Vogt et al, but not with the TF evolution found by Rix et al.  and
Simard \& Pritchet.  It is likely that some combination of size and
luminosity evolution is taking place, probably in different
proportions for galaxies of different luminosities and morphological
types.

\section{Conclusions} 
\label{sec:conc} 

Galaxy--galaxy lensing is a powerful tool for probing the structure of
dark matter halos at intermediate redshift.  In this paper, we have
detected the lensing signal due to the dark matter halos of galaxies
at the 99.3\% confidence level.  An important feature of our analysis
is the use of photometric redshifts to separate foreground and
background galaxies, to determine the relative distances between them,
and to obtain absolute magnitudes of the lens galaxies.

The scaling between lens galaxy halo circular velocity and galaxy $B$
luminosity is consistent with a TF relation with the same slope $\eta
= 0.35$ (or $7.14$ in magnitudes per log circular velocity) as is
observed locally. The typical lens galaxy, at a redshift $z = 0.6$,
has a circular velocity of $210\pm40 \kms$ at a rest-frame $B$
magnitude $M = -18.5$.

A number of control tests, in which lens and source positions and
source ellipticities are randomized, confirm the significance level of
the detection quoted above.  Furthermore, a marginal signal is also
detected from an independent, fainter sample of source galaxies
without photometric redshifts.  If the two source samples are
considered together, the lensing detection is significant at the
99.8\% confidence level.  Potential systematic biases, such as
contamination by satellite galaxies, distortion of the source shapes
by the light of the foreground galaxies, PSF anisotropies, and
contributions from mass distributed on the scale of galaxy groups and
on larger scales, were examined and are shown to be negligible.

Our HDF lens sample includes all morphological types but is dominated
by spiral galaxies $\sim 1$ mag.\ below $L_*$.  By comparing the
lensing $B$-band TF relation to that of local spirals, we find that at
a fixed circular velocity, the intermediate redshift galaxies are
somewhat fainter (by $1.0\pm0.6$ $B$ mag) than they are nearby.  This
result is consistent with the study of rotation curves of
intermediate-redshift spirals by Vogt et al.
The lack of brightening of individual galaxies, coupled with the
strong evolution in the global luminosity density suggests that the
latter is dominated by evolution in the galaxy number density.

The sizes of the dark matter halos are less well constrained than the
circular velocity due to the small size of the field.  Indeed, almost
any truncated isothermal model a mass of $2.8\pm1.2\times10^{11}
h^{-1} M_{\sun}$ within 30 \hkpc\ is an acceptable solution.  The
prospects for galaxy--galaxy lensing based on wide-field multi-color
imaging data are extremely promising: the wide ranges of separations
should allow a determination of the halo dark matter density in the
outer regions at $\sim 100\,\hkpc$.  On these scales the dark matter
halos of individual galaxies may merge into a common group or cluster
halo.  The problem of determining where dark matter halos end may
consequently prove to be more complicated, but no less important, than
was originally anticipated.

\acknowledgements

We thank the referee for useful suggestions which improved this paper.
MJH acknowledges financial support from a CITA National Fellowship and
from NSERC of Canada through operating grants to F.~D.~A. Hartwick and
C.~J. Pritchet.  SDJG gratefully acknowledges financial support from
his supervisor F.~D.~A.\ Hartwick.  HD acknowledges support from a PhD
research grant provided by the Research Council of Norway.

\end{document}